\documentstyle[12pt]{article}
\def\hybrid{\topmargin 0pt      \oddsidemargin 0pt
	\headheight 0pt \headsep 0pt
	\textheight 9in         
	\textwidth 6.25in       
	\marginparwidth .875in
	\parskip 5pt plus 1pt   \jot = 1.5ex}

\catcode`\@=11
\def\marginnote#1{}
\newcount\hour
\newcount\minute
\newtoks\amorpm
\hour=\time\divide\hour by60
\minute=\time{\multiply\hour by60 \global\advance\minute by-\hour}
\edef\standardtime{{\ifnum\hour<12 \global\amorpm={am}%
	\else\global\amorpm={pm}\advance\hour by-12 \fi
	\ifnum\hour=0 \hour=12 \fi
	\number\hour:\ifnum\minute<10 0\fi\number\minute\the\amorpm}}
\edef\militarytime{\number\hour:\ifnum\minute<10 0\fi\number\minute}

\def\draftlabel#1{{\@bsphack\if@filesw {\let\thepage\relax
   \xdef\@gtempa{\write\@auxout{\string
      \newlabel{#1}{{\@currentlabel}{\thepage}}}}}\@gtempa
   \if@nobreak \ifvmode\nobreak\fi\fi\fi\@esphack}
	\gdef\@eqnlabel{#1}}
\def\@eqnlabel{}
\def\@vacuum{}
\def\draftmarginnote#1{\marginpar{\raggedright\scriptsize\tt#1}}

\def\draft{\oddsidemargin -.5truein
	\def\@oddfoot{\sl preliminary draft \hfil
	\rm\thepage\hfil\sl\today\quad\militarytime}
	\let\@evenfoot\@oddfoot \overfullrule 3pt
	\let\label=\draftlabel
	\let\marginnote=\draftmarginnote
   \def\@eqnnum{(\theequation)\rlap{\kern\marginparsep\tt\@eqnlabel}%
\global\let\@eqnlabel\@vacuum}  }

\def\numberbysection{\@addtoreset{equation}{section}
	\def\theequation{\thesection.\arabic{equation}}}

\def\underline#1{\relax\ifmmode\@@underline#1\else
	$\@@underline{\hbox{#1}}$\relax\fi}

\def\titlepage{\@restonecolfalse\if@twocolumn\@restonecoltrue\onecolumn
     \else \newpage \fi \thispagestyle{empty}\c@page\z@
	\def\thefootnote{\fnsymbol{footnote}} }

\def\endtitlepage{\if@restonecol\twocolumn \else  \fi
	\def\thefootnote{\arabic{footnote}}
	\setcounter{footnote}{0}}  
\catcode`@=12
\relax

\def\beq{\begin{equation}}
\def\eeq{\end{equation}}
\def\bea{\begin{eqnarray}}
\def\eea{\end{eqnarray}}
\def\bar{\overline}

\def\nn{\nonumber}

\relax
\hybrid
\begin{document}
\begin{titlepage}
\setcounter{page}{0}
\begin{center}
 \hfill   PAR--LPTHE 92--4\\
 \hfill         January 1992\\[.4in]
{\large REMARKS ON THE PHYSICAL STATES AND THE CHIRAL
ALGEBRA OF 2D GRAVITY COUPLED TO $C \leq 1$ MATTER}\\[.4in]
	\large    Vl.S.Dotsenko\footnote{E-mail address:
	dotsenko@lpthe.jussieu.fr}\footnote{Permanent
address: Landau Institute for Theoretical
Physics, Moscow.}\\[.2in]
	{\it LPTHE\/}\footnote{Laboratoire associ\'e No. 280 au
CNRS}\\
       \it  Universit\'e Pierre et Marie Curie, PARIS VI\\
	Tour 16, 1$^{\it er}$ \'etage \\
	4 place Jussieu\\
	75252 Paris CEDEX 05, FRANCE\\

\end{center}

\vskip .3in
\centerline{ ABSTRACT}
\begin{quotation}
Some elaboration is given to the structure of physical states
in 2D gravity coupled to $C \leq 1$ matter, and to the chiral
algebra ($w_{\infty}$) of $C_{M} = 1$ theory which has been
found recently, in the continuum approach,
by Witten and by Klebanov and Polyakov. It is shown
then that the chiral algebra is being realized as well in
the minimal models of gravity ($C_{M}<1$), so that it stands
as a general symmetry of 2D gravity theories.

\end{quotation}
\end{titlepage}
\newpage

It can be said that the continuum field theory approach to
2D gravity has regained more attention recently. Two major
developments have been the complete classification of
the physical states in $C_{M} \leq 1$ gravity \cite{lz,bmp},
and the discovery of the ground ring and the chiral algebra
($w_{\infty}$) structures in $C_{M}=1$ theory \cite{w,kp}.
In the matrix model aproach related results have been obtained
in the recent papers \cite{aj}.

Here we shall first elaborate on the above results,
of the continuum theory approach,
and then show that the chiral algebra is present also in
the $C_{M}<1$ minimal models coupled to gravity. So that it
stands as a general symmetry of the theory of 2D gravity.
\vskip 1cm
The original BRST analysis of physical states in \cite{lz}
starts with the irreducible matter modules, the null states
being removed from the start. Then the physical states,
found in place of null states, involve ghost oscillators
and have nontrivial ghost numbers.

The second representation for extra physical states is
implicit in the analysis of \cite{bmp}, in which the matter
modules, realized as bosonic free field (Fock space) modules,
are made irreducible by using Felder's resolution \cite{f},
Fig.1. So, effectively, the null matter states are removed
from the start again. Then the extra physical states, of
the theory with gravity, appear as dressed tachyon type
states outside the basic grid. Now the ghost number grading
of the Lian and Zuckerman states is being replaced with
the Felder's resolution grading. The relation between
(the equivalence of) the two representations is studied
in the recent paper \cite{gjjm}.

We argue that there exists the third representation of
the minimal model extra states. One doesn't have to remove
the null states of the Fock space matter modules from
the start. There is no longer reason for doing that, as
the full theory, with gravity, is different. In particular,
the difinition of physical states is different. Then
the free field realized matter, together with the free field
realized Liouville, get extra states in the form of
nontrivially dressed (mixed with Liouville oscillators)
null states of the original matter theory.

Exercise could be made e.g. with the second level state
of the matter and Liouville common Fock space:
\beq
\Phi = [a_{1}(\partial\varphi_{M})^{2}+
        a_{2}\partial^{2}\varphi_{M}+
        b_{1}(\partial\varphi_{L})^{2}+
        b_{2}\partial^{2}\varphi_{L}+
        c\partial\varphi_{M}\partial\varphi_{L}]
\exp(i\alpha\varphi_{M})\exp(\beta\varphi_{L})
\label{L1}
\eeq
The physical state condition is
\bea
L_{n}\Phi &=& 0,\quad n>0
\label{L2}\\
L_{0}\Phi &=& \Phi
\label{L3}\\
\Phi &\neq& (k_{1}L_{-1}^{2}+k_{2}L_{-2})
\exp(i\alpha\varphi_{M})\exp(\beta\varphi_{L})
\label{L4}
\eea
($k_{1}, k_{2}$ stand for general coefficients). Here
$\{L_{n}\}$ are components of the full SET
\beq
T=T_{M}+T_{L},\quad T=\sum\frac{L_{n}}{z^{n+2}}
\label{L5}
\eeq
Realized in free fields, $T_{M}$ and $T_{L}$ have the standard
form for the theories with background charges:
\bea
T_{M} &=& -\frac{1}{4}(\partial\varphi_{M})^{2}
+i\alpha_{0}\partial^{2}\varphi_{M}, \quad
T_{L} = -\frac{1}{4}(\partial\varphi_{L})^{2}
+\beta_{0}\partial^{2}\varphi_{L}
\label{L6}\\
C_{M} &=& 1-24\alpha_{0}^{2}, \quad C_{L} = 1+24\beta_{0}^{2}
\label{L7}
\eea
with
\beq
C_{M}+C_{L}=26 \quad \longrightarrow \quad
\beta_{0}^{2}=1+\alpha_{0}^{2}
\label{L8}
\eeq
The two-point functions of $\varphi_{M}, \varphi_{L}$
in (\ref{L1}) and in (\ref{L6}) are normalized as
\beq
 \langle\varphi_{M}(z)\varphi_{M}(z')\rangle
=\langle\varphi_{L}(z)\varphi_{L}(z')\rangle
=2\log\frac{1}{z-z'}
\label{L9}
\eeq

By eq.(2) $\Phi$ has to be a primary field, and not
a descendent, by eq.(4).This is the old fashioned,
but basic definition for physical states in string
theory, and so in 2D gravity, which does not use
the representation with ghosts.

Requiring that the fourth and the third order pole
singularities did not appear in the operator product
expansion of $T(z), \Phi(z')$ (this is eq.(2)),
and that $\Phi$ were orthogonal to the states in
r.h.s. of eq.(4), one gets a linear system of five
equations. Condition that the solution exists fixes
the discrete values for the ``momenta'' $\alpha,
\beta$, to be either
\bea
\alpha&=&\alpha_{1.2}
\label{L10}\\
\beta&=&\beta_{1.2}
\label{L11}
\eea
or
\bea
\alpha&=&\tilde{\alpha}_{1.2} \equiv 2\alpha_{0}-\alpha_{1.2}
\label{L12}\\
\beta&=&\tilde{\beta}_{1.2} \equiv 2\beta_{0}-\beta_{1.2}
\label{L13}
\eea
(The notations for $\alpha_{n'.n}$, $\beta_{n'.n}$ are given
below, eqs. (19), (22), for genaral $\alpha_{0}$,
$\beta_{0}$ in this case, subject to eq.(8)).

Above is an example of a nontrivial dressing by
Liouville, which makes the would be null matter
state (on the second level, in the above example)
to become a physical state (stop being a descendant,
in particular) in the theory with gravity.

In general then the extra discrete physical states
get realized in this way all the way down inside
the original matter modules, those of the basic
grid only. This is indicated in Fig.2.

Existence of such states has been proved in \cite{bmp},
at the intermediate stage, in the proof of the theorem 3.
In the application of the theorem to the minimal model
the use of this fact was different, and has led to
the second representation of physical states, i.e.
the tachyon type states outside the basic grid, Fig.1.

The available calculations for the three-point
amplitudes involving extra states has so far been done
with the second representation \cite{d1,k}. We expect
naturally that all the representations should be equivalent,
and lead to same amplitudes, see also \cite{gjjm}.

In the limit of $C_{M} \rightarrow 1$ ($C_{L}
\rightarrow 25$) the basic grid becomes infinite.
The submodules structure becomes different, see
Fig.3. (Indicatively, the second top null state
in Fig.1 goes down to infinity). The submodules
get organized into the $su(2)$ multiplets, related to
each other by the action of the screening operators
\bea
H_{M}^{\pm}&=&\oint du \exp(i\alpha_{\pm}\varphi_{M}(u))
=\oint du \exp(\pm i \varphi_{M}(u))
\label{L14}\\
\alpha_{\pm}&=&\alpha_{0} \pm \sqrt{\alpha_{0}^{2}+1}=
\pm 1,\quad \hbox{for} \quad C_{M}=1 (\alpha_{0}=0)
\label{L15}
\eea

We remark that in $C_{M}=1$ theory the screening
operators have a different role. They define new states
- relate states to states (act locally). This is unlike
the $C_{M}=1$ theory where, in general, except for
the Felder's BRST operator $d'$, Fig.1, they relate
tensor products of states to states, Fig.4, defining
new, extra, operator product channels. It can be said
that in this second case the screening operators
represent (extra) properties of the vacuum, of
the theory with a background charge.

Existence of two moves between the Fock space
submodules, by the action of the operators
$H_{M}^{+}, H_{M}^{-}$, Fig.3, allows to prove that
the submodules are in fact decoupled - no inclusion
(with respect to the action of the Virasoro algebra
operators) but a direct sum. Then the top states of
the submodules already satisfy the eqs.(2) and (4) for
the physical states. Trivial dressing by a tachyonic
Liouville operator is sufficient to make them sutisfy
also the eq.(3), and make the discrete matter states
physical, with respect to $T=T_{M}+T_{L}$.

These facts about the $C_{M}=1$ theory are well known.
But let us remark on the corresponding modules of
$C_{L}=25$ Liouville theory. They are given in Fig.5.
In this particular case there is just one screening
operator, because for $C_{L}=25$
\beq
C=1+24\beta_{0}^{2}, \quad \beta_{0}=1 \label{L16}
\eeq
(The choice of $\beta_{0}=-1$ would be a trivial
redifinition of the theory, same e.g. as for
$\alpha_{0}$ , in $C<1$ matter theory).
\bea
\beta_{\pm}=\beta_{0} \pm \sqrt{\beta_{0}^{2}-1}=1
\label{L17}\\
H_{L}=\oint du \exp(\beta_{\pm} \varphi_{L}(u))
=\oint du \exp(\varphi_{L}(u)) \label{L18}
\eea
Accordingly, there is just one (regular) move between
the submodules, Fig.5. As a result, unlike the $C=1$
theory, the submodules are coupled inside the modules
(by the Virasoro operators) according to the inclusion
diagrams in Fig.5. There are no decoupled primary
states inside the modules, and so for dressing
the decoupled matter states one can use just the top
states of the modules - one Liouville tachyon type
(top) operator dressing the corresponding $su(2)$
multiplet of matter states. The second, across
the valley, top operator, of the same $(L_{0})_{Liouv}$
level, can also be used. It corresponds to the opposite
chirality, or the conjugate Coulomb gas representation
states.

The asymmetry between the $C=1$ and $C=25$ theories
(between the structures of bosonic, Fock space modules,
with respect to the corresponding Virasoro algebras)
indicates that the expected $su(2)$ structure of
the matter sector operator algebra is not going to get
cancelled fully by the Liouville, as it does not
possess the corresponding multiplets of decoupled
states.

The above remark (or worry) is due to the fact that in
the case of $C<1$ minimal theory the nontrivial operator
algebra (by which me imply here the three-point
amplitudes) of the matter does get cancelled by
Liouville, leaving just a product of normalization
factors \cite{gl,d1,k}. This can be thought of as to be
due to the complimentary structures of the corresponding
$C_{M}<1$ and $C_{L}>25$ theories, with respect to
the operators considered in those calculations.
\vskip 1cm
Next we shall elaborate somewhat on the calculation of
the chiral operator algebra by Klebanov and Polyakov
\cite{kp} for $C=1$ theory. We shall calculate
the general three-point amplitude for the discrete chiral
operators, without the assumption of $\mu=0$.

We shall first fix the notations. For the matter part
\beq
\alpha_{0}=0,\quad \alpha_{\pm}=\pm 1,\quad
\alpha_{n'.n}
=\frac{1-n'}{2}\alpha_{-}+\frac{1-n}{2}\alpha_{+}
=\frac{n'-n}{2}\equiv j
\label{L19}
\eeq
As $\alpha_{n'.n}$ depends on $(n'-n)$ in $C=1$ theory
we can use the parameter $j=(n'-n)/2$, which is
the $su(2)$ momentum with respect to
\bea
H_{M}^{\pm}&=& \oint \frac{du}{2\pi i}
\exp(\pm i\varphi_{M}(u))
\label{L20}\\
H_{M}^{0}&=& \oint \frac{du}{2\pi i}
\frac{i}{2}\partial\varphi_{M}(u)
\label{L21}
\eea
The tachyon type dressing of
$\exp(i\alpha_{n'.n}\varphi_{M})$ is by
$\exp(\beta_{-n'.n}\varphi_{L})$, or by
$\exp(\beta_{n'.-n}\varphi_{L})$, to ensure
$\Delta_{M}+\Delta_{L}=1$. Then for the Liouville part
\bea
\beta_{0}=1,\quad \beta_{\pm}=1,\quad
\beta_{-n'.n}
&=&\frac{1+n'}{2}\beta_{-}+\frac{1-n}{2}\beta_{+}
=1+\frac{n'-n}{2}=1+j
\label{L22}\\
\beta_{n'.-n}
&=&\frac{1-n'}{2}\beta_{-}+\frac{1+n}{2}\beta_{+}
=1-\frac{n'-n}{2}=1-j
\label{L23}
\eea

The conformal domensions of the operators are given by
\bea
\Delta_{n'.n}^{M}\equiv
\Delta(\exp(i\alpha_{n'.n}\varphi_{M}))
&=&\alpha_{n'.n}^{2}=j^{2}
\label{L24}\\
\Delta_{-n'.n}^{L}\equiv
\Delta(\exp(\beta_{-n'.n}\varphi_{L}))
&=&-(\beta_{-n'.n}^{2}-2\beta_{0}\beta_{-n'.n})=1-j^{2}
=\Delta_{n'.-n}^{L}
\label{L25}
\eea
They are indicated in Figs.3 and 5 for several first
integer $j$ operators.

The matter operators -
the multiplets of discrete states in Fig.3, are given by
\beq
\phi^{(M)j}_{\quad m}(z)=(H^{-}_{M})^{j-m}
\exp(ij\varphi_{M}(z))\equiv \prod^{j-m}_{1}\oint
\frac{du_{i}}{2\pi i}\exp(-i\varphi_{M}(u_{i}))
\exp(ij\varphi_{M}(z))
\label{L26}
\eeq
(The double use of $i$ in the r.h.s. is unlikely to be
confusing). Being dressed by the corresponding Liouville
operators they become spin one chiral operators of $C=1$
gravity \cite{w,kp}:
\beq
\Phi^{(\pm)j}_{\quad m}=\phi^{(M)j}_{\quad m}(z)
\phi^{(\pm,L)}_{j}(z)\equiv
\phi^{(M)j}_{\quad m}(z)\exp((1\mp j)\varphi_{L}(z))
\label{L27}
\eeq

The calculation of the three-point amplitudes
factorizes into the matter and the Liouville parts.
For the matter factor the $m$ dependence is absorbed,
due to the $su(2)$ symmetry, into the 3j symbols
of $su(2)$. Then we should expect:
\beq
\langle \phi^{(M)j_{1}}_{\quad m_{1}}(0)
\phi^{(M)j_{2}}_{\quad m_{2}}(1)
\phi^{(M)j_{3}}_{\quad m_{3}}(\infty)\rangle
\propto
(^{j_{1}\quad j_{2}
\quad j_{3}}_{m_{1}\quad m_{2}\quad m_{3}})
d^{(M)}_{j_{1}j_{2}j_{3}}
\label{L28}
\eeq
We can choose then the simpler but the general
$j_{1}j_{2}j_{3}$ case of
\bea
&&\langle
\phi^{(M)j_{1}}_{\quad -j_{1}}(0)
\phi^{(M)j_{2}}_{j_{1}-j_{2}}(1)
\phi^{(M)j_{3}}_{\quad j_{3}}(\infty)
\rangle\nn\\
&=&\langle \prod^{k}_{i=1} \oint_{C_{1(i)}}
\frac{du_{i}}{2\pi i} \exp(-i\varphi_{M}(u_{i}))
\exp(-ij_{1}\varphi_{M}(0))
\exp(ij_{2}\varphi_{M}(1))
\exp(ij_{3}\varphi_{M}(\infty))
\rangle\nn\\
&\propto & \prod^{k}_{i=1} \oint_{C_{1(i)}}
\frac{du_{i}}{2\pi i}
(u_{i})^{2j_{1}} (u_{i}-1)^{-2j_{2}} \prod^{k}_{i<j}
(u_{i}-u_{j})^{2}
\label{L29}
\eea
Here $k=-j_{1}+j_{2}+j_{3}$ (ought to be integer).
The contours {$C_{1(i)}$} encircle the point $z=1$
as in Fig.6(left). We shall use also the notation
$n_{1}=2j_{1}, n_{2}=2j_{2}, n_{3}=2j_{3}$
for the integer-valued exponents in (29). So we have
the integral
\beq
I^{(M)}_{k}(n_{1},-n_{2};1)=
\prod^{k}_{i=1} \oint_{C_{1(i)}}
\frac{du_{i}}{2\pi i}
(u_{i})^{n_{1}} (u_{i}-1)^{-n_{2}} \prod^{k}_{i<j}
(u_{i}-u_{j})^{2}
\label{L30}
\eeq
It can be calculated by transforming it to the standard
form in \cite{df}. First we move the contours in Fig.6
(left to right), Which can be done since $z=0$ is
a regular point in (30). And then we shift the exponents
by a small amount of $\epsilon$ (to regularize
the integral below):
\beq
n_{1}\rightarrow a=n_{1}+\epsilon,\quad
n_{2}\rightarrow b=-n_{2}-\epsilon
\label{L31}
\eeq
This results in the cut line between $z=0$ and $z=1$,
but not outside, so that the contours are still closed,
Fig.7. The resulting integral is standard. It is given
by \cite{df}:
\bea
I^{(M)}(a,b;1)&\propto&(\frac{sin\pi a}{\pi})^{k}
\prod^{k}_{i=1}\int^{1}_{0}du_{i}(u_{i})^{a}
(1-u_{i})^{b}\prod^{k}_{i<j}(u_{i}-u_{j})^{2}\nn\\
&\simeq& (\epsilon)^{k} k! \prod^{k}_{i=1}
\frac{\Gamma(i)}{\Gamma(1)}
\frac{\Gamma(1+a+i-1)\Gamma(1+b+i-1)}{\Gamma(2+a+b+k-2+i)}
\nn\\
&\simeq& (\epsilon)^{k} k! \prod^{k}_{i=1} \Gamma(i)
\frac{\Gamma(n_{1}+i) \Gamma(-n_{2}+i-\epsilon)}{\Gamma(n_{1}-n_{2}+k+i)}
\label{L32}
\eea
Next we use
\beq
\Gamma(-n_{2}+i-\epsilon)\simeq
\frac{(-1)^{n_{3}-i+1}}{\epsilon}
\frac{1}{\Gamma(n_{2}-i+1)}
\label{L33}
\eeq
to get (up to signs, which are generally ignored below)
\beq
I^{(M)} = k! \prod^{k}_{i=1}\frac{\Gamma(i)\Gamma(n_{1}
+i)}{\Gamma(n_{1}-n_{2}+k+i)\Gamma(n_{2}-i+1)}
\label{L34}
\eeq
Using the relations for the products like e.g.
\beq
\prod^{\frac{-n_{1}+n_{2}+n_{3}}{2}}_{i=1}
\Gamma(n_{1}+i)=
\prod^{\frac{n_{1}+n_{2}+n_{3}}{2}}_{i=n_{1}+1}
\Gamma(i)=
\prod^{n_{1}}_{i=1}\frac{1}{\Gamma(i)}
\prod^{\frac{n_{1}+n_{2}+n_{3}}{2}}_{i=1}
\Gamma(i)
\label{L35}
\eeq
we obtain from (34):
\bea
I^{(M)}&=&\frac{1}{(j_{1}+j_{2}-j_{3})!(j_{1}-j_{2}+j_{3})!}
(\Delta_{j_{1}j_{2}j_{3}})^{1/2}((2j_{1})!(2j_{2})!(2j_{3})!)^{1/2}
\times d^{(M)}_{j_{1}j_{2}j_{3}}
\nn\\
d^{(M)}_{j_{1}j_{2}j_{3}}&=&P(j_{1}+j_{2}+j_{3})
P(-j_{1}+j_{2}+j_{3}+1)
P(j_{1}-j_{2}+j_{3}+1)
P(j_{1}+j_{2}-j_{3}+1)
\nn\\
&\times&
P^{-1}(2j_{1})
P^{-1}(2j_{2})
P^{-1}(2j_{3})
\times (\Delta_{j_{1}j_{2}j_{3}})^{-1/2}
((2j_{1})!(2j_{2})!(2j_{3})!)^{-1/2}
\label{L36}
\eea
Here
\bea
&&P(n)=\prod^{n}_{i=1}\Gamma(i)\nn\\
&&\Delta_{j_{1}j_{2}j_{3}}=
\frac{(j_{1}+j_{2}-j_{3})!(j_{1}-j_{2}+j_{3})!(-j_{1}+j_{2}+
j_{3})!}{(j_{1}+j_{2}+j_{3}+1)!}
\label{L37}
\eea
For the specific normalization of the $su(2)$ operators
which is used here (the operators (20), (21), (26)):
\beq
\langle H^{+}\exp(-ij\varphi(0))\times H^{-}\exp(ij\varphi(1))
\rangle = (2j)!\frac{(j-m)!}{(j+m)!}
\label{L38}
\eeq
the first factor in (36) is the 3j symbol
\beq
(^{j_{1}\quad \quad j_{2}\quad \quad j_{3}}_{-j_{1}\quad j_{1}-j_{2}
\quad j_{3}})=
\frac{1}{(j_{1}+j_{2}-j_{3})!(j_{1}-j_{2}+j_{3})!}
(\Delta_{j_{1}j_{2}j_{3}})^{1/2}((2j_{1})!(2j_{2})!(2j_{3})!)^{1/2}
\label{L39}
\eeq

The three-point function for the general case is then
given by eq.(28), with the coefficients
$d^{(M)}_{j_{1}j_{2}j_{3}}$ given by eq.(36).
We can assume then the standard normalization
of the $su(2)$ representation operators, by a unit.
Then the $su(2)$ 3j symbols in eq.(28) become the standard
ones.

We remark that the three-point functions (28),(36)
for the operators $\phi^{(M)j}_{\quad m}(z)$, eq.(26),
are different from the Wess-Zumino $su(2)$ theory
operator algebra structure constants, calculated in
\cite{zf}, because $\phi^{(M)j}_{\quad m}(z)$ are not
the primary operators, with respect to the currents:
\beq
J^{\pm}(z)=\exp(\pm i\varphi_{M}(z)),\quad
J^{0}(z)=\frac{i}{2}\partial \varphi_{M}(z)
\label{L40}
\eeq
(comp. eqs. (20),(21)). The above currents are known
to form the level one $su(2)$ current algebra \cite{fk}.

On the Liouville side of the three-point amplitude
we shall first choose the representation
\beq
\langle
\exp((1-j_{1})\varphi_{L}(0))
\exp((1-j_{2})\varphi_{L}(1))
\exp((1+j_{3})\varphi_{L}(\infty))
\rangle
\label{L41}
\eeq
i.e. the one with the ``signature'' (+ + -), comp.
eq.(27). Without the screening operators the momentum
conservation (anomalous,due to the background charge)
is given by:
\bea
\sum\beta_{i}&=&2\beta_{0} \quad \rightarrow \quad
3-j_{1}-j_{2}+j_{3}=2\nn\\
j_{3}&=&j_{1}+j_{2}-1
\label{L42}
\eea
With the Liouville screening operators added,
$\mu\int du \exp(\varphi(u))$, The Liouville part of
the three-point amplitude becomes:
\beq
\frac{1}{(l)!}(\mu)^{l}\langle
\exp((1-j_{1})\varphi_{L}(0))
\exp((1-j_{2})\varphi_{L}(1))
\exp((1+j_{3})\varphi_{L}(\infty))
\prod^{l}_{i=1}\int^{+\infty}_{-\infty} du_{i}
\exp(\varphi(u_{i}))\rangle
\label{L43}
\eeq
The momentum conservation takes the general form,
for the theory with the background charge and
the screening operators in the vacuum:
\bea
\sum\beta_{i}+l\beta_{-}&=&2\beta_{0} \quad \rightarrow
\quad 3-j_{1}-j_{2}+j_{3}+l=2 \nn\\
j_{3}&=&j_{1}+j_{2}-1-l
\label{L44}
\eea
Particular $l$ picks up a particular channel of
the operator product algebra (OPA) - the particular value of
$j_{3}$. (So far this is parallel to the minimal model
setting of the OPA).

Taking the average, the eq.(43) becomes:
\beq
I^{(L)}_{l} \propto \frac{1}{l!}(\mu)^{l}\prod^{l}_{i=1}
\int^{+\infty}_{-\infty}du_{i}(u_{i})^{-2(1-j_{1})}
(u_{i}-1)^{-2(1-j_{2})}\prod^{l}_{i<j}(u_{i}-u_{j})^{-2}
\label{L45}
\eeq

We have to decide on the choice of contours of
integration in (45). The apparent ambiguity is due to
the chiral calculation of the three-point function.
We shall assume for the moment, for the purpose of
the discussion of the contours, that all three points
are at a finite position  on the complex plane:
\beq
I^{(L)}_{l} \propto \frac{1}{l!}(\mu)^{l}\prod^{l}_{i=1}
\int^{+\infty}_{-\infty}du_{i}(u_{i}-z_{1})^{2(j_{1}-1)}
(u_{i}-z_{2})^{2(j_{2}-1)}
(u_{i}-z_{3})^{-2(j_{1}+j_{2}-l)}
\prod^{l}_{i<j}(u_{i}-u_{j})^{-2}
\label{L46}
\eeq
We have used eq.(44), by which the exponent of
$(u_{i}-z_{3})$ is $-2(1+j_{3})=-2(j_{1}+j_{2}-l)$.

The screening operators could be understood as coming
from the expansion over the cosmological (exponential)
term in the Liouville action. We could think of
the $u$ - contour integrals as coming from
the functional integral representation, with
the 2D integration over the cosmological operators being
transformed into a sum of products of $u$ and $\bar{u}$
contour integrals. The description of this standard
analytic technique is given e.g. in \cite{d}.
Then, whatever the absorption of the $\bar{u}$ factors,
on the $u$ chiral side just the contours in Fig.8 would
appear. To simplify pictures we consider just one
contour integration. The present (heuristic basically)
discussion could be extended to the multiple contour
case as well.

The cases $I$ and $IV$ give vanishing result, whatever
are the singularities at the points $z_{1}$, $z_{2}$,
$z_{3}$. The contour could be shifted in these cases
to infinity, where the integral converges and so
vanishes over a shrinking loop. There remain
configurations $II$ and $III$, which could also be
transformed to the configurations $II_{1}$ or $II_{2}$,
and $III_{1}$ or $III_{2}$ respectively, Fig.9.
Then, if $2(j_{1}-1)\geq 0$, $2(j_{2}-1)\geq 0$,
the integral $I^{L}$ vanishes, for either contour,
II or III, as the points $z_{1}$ and $z_{2}$ are
regular, and the contours in $II_{2}$ or $III_{1}$
shrink to zero. Same happens with the multiple
contour case, see Fig.10. Still this vanishing
might be misleading, as in $C=1$ theory one assumes
usually a singular renormalization of the cosmological
constant $\mu$, by $1/\Gamma(0)$.

In the other treatment, by taking e.g.
the signature (- - +), comp. eq.(27), we shall get
\bea
I^{(L)}_{l}\propto
&&\frac{1}{(l)!}(\mu)^{l}
\langle
\exp((1+j_{1})\varphi_{L}(z_{1}))
\exp((1+j_{2})\varphi_{L}(z_{2}))\nn\\
&&\times \exp((1-j_{3})\varphi_{L}(z_{3}))
\prod^{l}_{i=1}\int^{+\infty}_{-\infty} du_{i}
\exp(\varphi(u_{i}))
\rangle\nn\\
&&\propto \frac{1}{(l)!}(\mu)^{l}
\prod^{l}_{i=1}
\int^{+\infty}_{-\infty}du_{i}(u_{i}-z_{1})^{-2(j_{1}+1)}
(u_{i}-z_{2})^{-2(j_{2}+1)}\nn\\
&&\times (u_{i}-z_{3})^{2(j_{1}+j_{2}-l)}
\prod^{l}_{i<j}(u_{i}-u_{j})^{-2}
\label{L47}
\eea

The nontrivial
contours in this case are those in Fig.11 (assuming
$2(j_{1}+j_{2}-l)\geq 0$). The calculation for this
case, which uses the analytic continuation tricks,
see the Appendix, leads to the result:
\bea
I^{(L)}_{l}\propto &&(\frac{\mu}{\Gamma(-1)})^{l}
\frac{1}{\Gamma(0)}
P^{-1}(j_{1}+j_{2}+j_{3})\nn\\
&&\times P^{-1}(-j_{1}+j_{2}+j_{3})
P^{-1}(j_{1}-j_{2}+j_{3})
P^{-1}(j_{1}+j_{2}-j_{3})\nn\\
&&\times P(2j_{1}+1)P(2j_{2}+1)P(2j_{3}-1)
\label{L48}
\eea
The factor $1/\Gamma(0)$, though singular, is
the standard overall normalization of the Liouville
calculation of the amplitudes, see \cite{d1}, which
would disappear if the amplitude is normalized
by the partition function.

Combining eq.(28) and eq.(48) we find:
\bea
&&\langle
\Phi^{(-)j_{1}}_{\quad m_{1}}
\Phi^{(-)j_{2}}_{\quad m_{2}}
\Phi^{(+)j_{3}}_{\quad m_{3}}
\rangle
\sim
\frac{1}{\Gamma(0)}(\frac{\mu}{\Gamma(-1)})^{l}\nn\\
&&\times (^{j_{1}\quad j_{2}
\quad j_{3}}_{m_{1}\quad m_{2}\quad m_{3}})
\times (\Delta_{j_{1}j_{2}j_{3}})^{-1/2}
((2j_{1})!(2j_{2})!(2j_{3})!)^{-1/2}\nn\\
&&\times (j_{1}+j_{2}-j_{3})!(j_{!}-j_{2}+j_{3})!
(-j_{1}+j_{2}+j_{3})!
\times (2j_{1})!(2j_{2})!\frac{1}{(j_{3}-1)!}
\label{L49}
\eea

For $j_{3}=j_{1}+j_{2}-1$ one gets the result of \cite{kp}
(in representation used above this corresponds to
$l=-2$, see eq.(67) in the Appendix):
\bea
&&\langle
\Phi^{(-)j_{1}}_{\quad m_{1}}
\Phi^{(-)j_{2}}_{\quad m_{2}}
\Phi^{(+)j_{1}+j_{2}-1}_{\quad m_{3}}
\rangle
\sim
(j_{2}m_{1}-j_{1}m_{2})\times
\frac{N(j_{1},m_{1})N(j_{2},m_{2})}{N(j_{3},m_{3})}\nn\\
&&N(j,m)=\frac{((2j)!)^{1/2}(2j-1)!}{((j+m)!(j-m)!)^{1/2}}
\label{L50}
\eea
We have dropped here the singular factors in front.

The analysis of the general three-point amplitude,
eq.(49), with respect to its chiral algebra
interpretation, will be considered elsewhere.

\vskip 1cm
Let us consider now the $C_{M}<1$ minimal matter
coupled to gravity. We want to show that
the $w_{\infty}$ chiral algebra is present
in this theory as well.

We take now the matter and the Liouville SETs
(for this theories being represented by free fields
$\varphi_{M}$, $\varphi_{L}$) in the form:
\bea
&&T_{M}=-\frac{1}{2}(\partial\varphi_{M})^{2}
+i\alpha_{0}\partial^{2}\varphi_{M},\quad
T_{L}=-\frac{1}{2}(\partial\varphi_{L})^{2}
+\beta_{0}\partial^{2}\varphi_{L}
\nn\\
&&C_{M}=1-12\alpha_{0}^{2},\quad C_{L}=1+12\beta_{0}^{2}
\nn\\
&&C_{M}+C_{L}=26 \quad \rightarrow \quad
\beta_{0}^{2}=2-\alpha_{0}^{2}
\label{L51}
\eea
This corresponds to the normalization of the two-point
functions
\beq
\langle\varphi_{M}(z)\varphi_{M}(z')\rangle=
\langle\varphi_{L}(z)\varphi_{L}(z')\rangle=
\log\frac{1}{z-z'}
\label{L52}
\eeq
being changed as compared to eqs.(6)-(9).
(These changes of normalization are to avoid
$\sqrt{2}$ in the formulas). The full SET of the theory
is:
\bea
&&T=T_{M}+T_{L}\nn\\
&&=-\frac{1}{2}
((\partial\varphi_{M})^{2}
+(\partial\varphi_{L})^{2})
+i\alpha_{0}\partial^{2}\varphi_{M}
+\beta_{0}\partial^{2}\varphi_{L}
\label{L53}
\eea

Let us make the rotation in the space of
($\varphi_{M}$, $\varphi_{L}$) of the form:
\bea
\varphi_{1}=-i\alpha_{0}\varphi_{M}
-\beta_{0}\varphi_{L}\nn\\
\varphi_{2}=\beta_{0}\varphi_{M}
-i\alpha_{0}\varphi_{L}
\label{L54}
\eea
or
\bea
\varphi_{M}=\frac{1}{2}(-i\alpha_{0}\varphi_{1}
+\beta_{0}\varphi_{2})\nn\\
\varphi_{L}=\frac{1}{2}(-\beta_{0}\varphi_{1}
-i\alpha_{0}\varphi_{2})
\label{L55}
\eea
Then
\bea
(\partial\varphi_{M})^{2}+(\partial\varphi_{L})^{2}
=\frac{1}{2}
((\partial\varphi_{1})^{2}+(\partial\varphi_{2})^{2})
\nn\\
i\alpha_{0}\partial^{2}\varphi_{M}
+\beta_{0}\partial^{2}\varphi_{L}
=-\partial^{2}\varphi_{1}
\label{L56}
\eea
and we find that $T$ in eq.(53 takes the form:
\beq
T=-\frac{1}{2}(\partial\varphi_{2})^{2}
-\frac{1}{2}(\partial\varphi_{1})^{2}
-\partial^{2}\varphi_{1} \equiv T_{2}+T_{1}
\label{L57}
\eeq

SET $T_{2}$ is that of (matter$)_{2}$ with $C_{2}=1$,
no background charge, and $T_{1}$ is that of
(Liouville$)_{1}$, with the background charge
\bea
(\beta_{0})_{1}=-1
\label{L58}\\
C_{2}=1+24\beta_{0}^{2}=25
\label{L59}
\eea
Essential for this rotation to be an allowed
transformation is that $\varphi_{M}$ and $\varphi_{L}$
to be treated on equal footing, as free fields,
with background charges.

The operators of the theory are:
\bea
\Phi^{-}_{n'.n}=\exp(i\alpha_{n'.n}\varphi_{M}
+\beta_{-n'.n}\varphi_{L})
\label{L60}\\
\Phi^{+}_{n'.n}=\exp(i\alpha_{n'.n}\varphi_{M}
+\beta_{n'.-n}\varphi_{L})
\label{L61}
\eea
In the new basis of ($\varphi_{1}$, $\varphi_{2}$)
the operators $\Phi^{-}_{n'.n}$ take the form:
\bea
\Phi^{-}_{n'.n}=\exp(ij_{n'.n}\varphi_{2}
-(1+j_{n'.n})\varphi_{1})
\label{L62}\\
j_{n'.n}=\frac{\rho'n'-n}{2}, \quad
\rho'=\frac{(\alpha_{-})^{2}}{2}=\frac{p}{p'}
\label{L63}
\eea
Here $p$, $p'$ are the usual parameters of the minimal
conformal theory. For the border case operators
$n'=0$, see Fig.12, we shall have the operators (62)
with
\beq
j_{0.n}=-\frac{n}{2}
\label{L64}
\eeq
(Shifts of $n'$ by $k\times p'$, for $k$ integer, result in the same
set of operators). They have the structure of (matter$)_{2}$ and
(Liouville$)_{1}$ modules and submodules in Fig.3
and in Fig.5, which has been discussed in the first
part of the paper. The corresponding extra discrete
operators, the chiral ones, will form the chiral algebra
$w_{\infty}$.

The rest - the ``regular'' operators in (62), those with
$n'\neq kp'$ ($k$ is integer), are tachyonic operators
of $C=1$ gravity, with the (matter$)_{2}$ momentum
taking a discrete set of values in eq.(63). These are
general position operators, in a sence that their
(matter$)_{2}$ modules are not degenerate, and involve
no extra decoupled discrete states (if considered within
the matter theory alone).

The operators $\Phi^{+}_{n'.n}$, eq.(61), might be
giving an alternative representation for
 the same
structure. In particular, the chiral operators
would correspond to $n=0$, up to $k\times p$ shifts.
This is not yet clear, though.

The linear transformation in the space of
$\partial\varphi_{M}$, $\partial\varphi_{L}$, to
the light-cone like oscillator basis of
$\partial\varphi^{\pm} \sim \partial\varphi_{M}
\pm\partial\varphi_{L}$ is used in the papers
\cite{lz,bmp}, in  the study of BRST cohomology, i.e.
the physical states spectrum of 2D gravity.
The rotation in the space of $\varphi_{M}$, $\varphi_{L}$
have also been discussed by Polyakov \cite{p}.

\noindent{\large Acknowledgements}

I am grateful to the colleagues at LPTHE for their
kind hospitality. I have benefited from discussions
with L.Baulieu, V.A.Fateev, B.Feigin, G.Felder, E.Frenkel,
J.-L.Gervais, E.Kiritsis, M.Lashkevich, A.M.Polyakov,
P.Windey.

\newpage


APPENDIX

The integral (47), for $z_{1} \rightarrow \infty$,
$z_{2}=0$, $z_{3}=1$, and the contour integration on
Fig.11, takes the form (we drop the $(\mu)^{l}$ factor
here):
\beq
I^{(L)}_{l} \propto \frac{1}{l!}(\mu)^{l}\prod^{l}_{i=1}
\oint_{C_{0,1}}du_{i}(u_{i})^{-2(j_{2}+1)}
(u_{i}-1)^{2(j_{1}+j_{2}+l)}\prod^{l}_{i<j}(u_{i}-u_{j})^{-2}
\label{L65}
\eeq
It is of the standard form \cite{df}, and ane gets
(comp. also eqs.(30),(32), for $I^{M}$):
\bea
&&I^{(L)}\propto (\frac{\sin 2\pi(j_{2}+1)}{\pi})^{l}\nn\\
&&\times
\prod_{i=1}^{l}\frac{\Gamma(-i)}{\Gamma(-1)}
\frac{\Gamma(1-2(1+j_{2})-i+1)\Gamma(1+2(j_{1}+j_{2}+l)-i+
1)}{\Gamma(2-2(1+j_{2})+2(j_{1}+j_{2}+l)-(l-2+i))}\nn\\
&&\propto\frac{1}{(\Gamma(-1))^{l}}(\frac{\pi}{\sin\pi(n_{2}+2)})^{k+1}
\prod_{i=0}^{k}
\frac{\Gamma(-n_{2}+n_{3}+k+1+i)}{\Gamma(i)\Gamma(-n_{2}+i)\Gamma(n_{3}+i)}
\label{L66}
\eea
Here $n_{1}=2j_{1}$, $n_{2}=2j_{2}$, $n_{3}=2j_{3}$,
\bea
&&l=-j_{1}-j_{2}+j_{3}-1,\quad j_{1}+j_{2}+l=j_{3}-1
\label{L67}\\
&&k=|l|-1=j_{1}+j_{2}-j_{3}=\frac{n_{1}+n_{2}-n_{3}}{2}
\label{L68}
\eea
Here we have assumed $l$ to be negative, $k$ to be
positive, and have used the analytic continuation
trick for the products \cite{d1}:
\beq
\prod_{i=1}^{l} f(i)=\prod_{i=0}^{|l|-1}\frac{1}{f(-i)}
\label{L69}
\eeq

For handling the expression in (66) we shall also use
the relation, for $\Gamma(-n_{2}+i)$,
\beq
\Gamma(-n)=\frac{\Gamma(0)}{\Gamma(n+1)} (-1)^{n}
\label{L70}
\eeq
Then, dropping the sign factors throughtout, we shall
get for (66):
\bea
&&I^{(L)}\propto\frac{1}{(\Gamma(-1))^{l}} \frac{1}{(0)^{k+1}}
\prod_{i=0}^{k}
\frac{\Gamma(n_{2}+1-i)\Gamma(-n_{2}+n_{3}+k+1+
i)}{\Gamma(i)\Gamma(n_{3}+i)\Gamma(0)}
\nn\\
&&=\frac{1}{(\Gamma(-1))^{l}} \prod_{i=0}^{k}
\frac{\Gamma(n_{2}+1-i)\Gamma(-n_{2}+n_{3}+k+1+
i)}{\Gamma(i)\Gamma(n_{3}+i)}
\label{L71}
\eea
The treatment of the products is similar to that in
the case of the integral $I^{(M)}$, eq.(32), but with one
extra factor of $1/\Gamma(0)$:
\bea
&&\prod_{i=1}^{k}\frac{1}{\Gamma(i)}=\frac{1}{\Gamma(0)}
P^{-1}(j_{1}+j_{2}-j_{3})
\label{L72}\\
&&\prod_{i=0}^{k}\frac{1}{\Gamma(n_{3}+i)}
=\prod_{i=0}^{\frac{n_{1}+n_{2}-n_{3}}{2}}\frac{1}{\Gamma(n_{3}+i)}
=\prod_{i=n_{3}}^{\frac{n_{1}+n_{2}+n_{3}}{2}}\frac{1}{\Gamma(i)}
\nn\\
&&=P^{-1}(j_{1}+j_{2}+j_{3})P(2j_{3}-1)
\label{L73}\\
&&\prod_{i=0}^{k}\Gamma(-n_{2}+n_{3}+k+1+i)=
\prod_{i=0}^{\frac{n_{1}+n_{2}+n_{3}}{2}}
\Gamma(\frac{n_{1}+n_{2}+n_{3}}{2}+1+i)=
\prod_{i=\frac{n_{1}+n_{2}+n_{3}}{2}+1}^{n_{1}+1}
\Gamma(i)\nn\\
&&=P^{-1}(j_{1}-j_{2}+j_{3})P(2j_{1}+1)
\label{L74}\\
&&\prod_{i=0}^{\frac{n_{1}+n_{2}+n_{3}}{2}}
\Gamma(n_{2}+1-i)=
\prod_{i=\frac{-n_{1}+n_{2}+n_{3}}{2}+1}^{n_{2}+1}
\Gamma(i)\nn\\
&&=P^{-1}(-j_{1}+j_{2}+j_{3})P(2j_{2}+1)
\label{L75}
\eea
Finally one gets, with $(\mu)^{l}$ placed back,
\bea
&&I^{(L)}\propto (\frac{\mu}{\Gamma(-1)})^{l}
\frac{1}{\Gamma(0)}\nn\\
&&\times
P^{-1}(j_{1}+j_{2}+j_{3})
P^{-1}(-j_{1}+j_{2}+j_{3})
P^{-1}(j_{1}-j_{2}+j_{3})
P^{-1}(j_{1}+j_{2}-j_{3})
\nn\\
&&\times
P(2j_{1}+1)
P(2j_{2}+1)
P(2j_{3}-1)
\label{L76}
\eea

\newpage


\newpage


FIGURE CAPTIONS

Fig.1. Shadowed circles indicate the discrete (infinite
set of states that remain, as physical, in the theory
with gravity. Shown is only one set of states, which is
generated by the primary field $\Phi_{2.2}$. Similar
sets of states are associated with each primary field
in the basic grid of the original matter theory.

Explicit form of the ``moves'', or differentials $d'$
is realized by
specific multiple integrals of the screening operators,
see \cite{f}.

Fig.2. The suggested third representation of the physical
states, as null states of matter (tops of the submodules)
nontrivially dressed with Liouville. Just the $\Phi_{2.2}$
set of states is indicated. Similar seta are associated
with each primary field inside the basic grid of
the original matter theory.

Fig.3. Set of matter $C_{M}=1$ modules, for $j$ integer.
On the r.h.s. the inclusion diagram of the submodules,
for a particular module, is indicated. It is a direct
sum in this case.

Fig.4. Different role of the screening operators
in $C=1$ and in $C<1$ theories.

Fig.5. The set of Liouville $C_{L}=25$ modules,
for $j$ integer. On the left and on the right
are given the inclusion diagrams, for two particular
modules.

Fig.6. The contours of the matter three-point function

Fig.7. The contours of the matter three-point function,
with $\epsilon$ -regularized exponents.

Fig.8. Four distinct types of the $+\infty$ to $-\infty$
contours for the Liouville three-point function.

Fig.9. The closed contour integrals corresponding
to the type II and III contours in Fig.8.

Fig.10. The transformation between
the $+\infty$ to $-\infty$ and closed contours,
for the multiple contour case.

Fig.11. The closed contours for the Liouville
three-point function integral in eq.(65).

\end{document}